\documentclass{aa} 
\bibpunct{(}{)}{;}{a}{}{,} 
\usepackage{graphicx}
\usepackage{txfonts}
\begin{document} 
\nolinenumbers

   \title{Beryllium: The smoking gun of a rejuvenated star}

   \author{A. Rathsam\inst{1}
          \and
          J. Mel\'{e}ndez\inst{1}
          \and
          A. I. Karakas\inst{2,3}
          }

   \institute{Universidade de São Paulo, Instituto de Astronomia, Geofísica e Ciências Atmosféricas, IAG, Departamento de Astronomia, Rua do Matão 1226, Cidade Universitária, 05508-090, SP, Brazil
             \and
              School of Physics and Astronomy, Monash University, Clayton, VIC 3800, Australia
              \and
              ARC Centre of Excellence for Astrophysics in Three Dimensions (ASTRO-3D), Melbourne, VIC 3000, Australia \\
              \email{annerathsam@usp.br, jorge.melendez@iag.usp.br}
             }

    %
    %

   \date{Received XXX; accepted YYY}

 
  \abstract
   {The chemistry and Galactic velocity components of the star HD 65907 suggest that despite its young isochronal age of $\sim$5 Gyr, it is in fact a merger of two old Population II stars. Its low Li abundance is also consistent with a mass accretion episode.}
   {We determine Li and Be abundances for this star and evaluate its radial velocity time series, activity cycle, and spectral energy distribution in search of clues regarding the origin of this enigmatic star.}
   {Li and Be abundances were determined via spectral synthesis of their resonance lines using HARPS and UVES spectra, respectively. HARPS data were also used to study variations in the star's radial velocity and activity levels. Photometric data were adopted to evaluate the stellar spectral energy distribution.}
   {HD 65908 is severely Li- and Be-depleted. Its radial velocity is nearly constant ($\sigma =$ 2 m/s), with a small modulation likely associated with stellar activity, and the star shows no further signs of an undetected close companion. The excess infrared emission is consistent with a 30 K blackbody, which is interpreted as a debris disk surrounding the star. The post-merger mass, rotation rate, and evolution of this star  are discussed.}
   {The low Li and Be abundances, in addition to the lack of evidence for a companion, are strong pieces of evidence in favor of the stellar merger scenario. In this context, Be can be used to confirm other blue stragglers among field solar-type stars, as proposed in the literature.}

   \keywords{stars: abundances --
                blue stragglers --
                stars: individual (HD 65907) --
                stars: Population II
               }

   \maketitle
%
\nolinenumbers
\section{Introduction}

    Blue stragglers are stars that are bluer and more luminous than expected, meaning that they appear younger than their actual age. They are usually spotted in stellar clusters \citep{sandage}, where they have effective temperatures and luminosities higher than the turnoff point for the population to which belong. The identification of field blue stragglers, although possible, is more difficult since we do not usually know their true age a priori. For these stars, we must take advantage of different age indicators as we look for clues regarding the real stellar age. In this sense, the star HD 65907 (HIP 38908/HR 3138) presents many signs that it is a merger of two old Pop. II stars, despite its isochronal age of about 5 Gyr (5.6$^{+1.1}_{-0.9}$ Gyr according to \citealt{fuhrmann} and 4.60$^{+0.53}_{-0.69}$ Gyr according to \citealt{shejeela}; all uncertainties are within 1$\sigma$).

    \citet{fuhrmann} first noticed that the chemical composition and kinematics of this star are not compatible with what is expected for a thin disk star. HD 65907 has a high [Mg/Fe] ($+0.34$ dex $\pm$ 0.05; \citealt{fuhrmann}), which is typical of the thick disk population, due to the fact that during the formation of this structure, the interstellar medium was enriched by type II supernova contamination but not yet by type Ia supernovae (\citealt{haywood, gce_prantzos, gce_kobayashi}). Additionally, this star has high U (toward the Galactic center) and W (perpendicular to the Galactic plane) velocity components, which is again typical of old Pop. II stars. \citet{shejeela} also note that this star has [Y/Mg] = -0.32 $\pm$ 0.01 dex; this implies a chemical age of $\sim11$ Gyr, which differs significantly from its isochrone-derived age.

    The stellar chemistry holds other clues that HD 65907 underwent some puzzling phenomena. HD 65907 is a subsolar metallicity F-type star ([Fe/H] = -0.315 $\pm$ 0.005 dex, M = 1.02$^{+0.01}_{-0.02}$ M$_\odot$; \citealt{shejeela}), which are stars with shallow convection zones. This implies that a large destruction of Li is not expected, although main-sequence stars continuously burn Li at a low rate throughout their evolution (e.g., \citealt{marilia, dumont}; \citealt{eu}). However, this star has low Li abundances, with the lowest values reported by the most recent works, which are based on spectra of higher quality: log(N(Li)) $<$ 0.9 dex \citep{li_87}, log($\epsilon$(Li)) $=$ 0.78 to 1.20 dex \citep{li1}, log($\epsilon$(Li)) $<$ 0.98 dex \citep{li2}, A(Li) $<$ 0.73 dex \citep{li3}, and A(Li) $<$ 0.15 dex \citep{li4}. These low Li abundances cannot be explained regardless of whether we consider it a Pop. I or II star -- a low-metallicity, 5 Gyr old, F-type Pop. I star should not have such low Li abundances \citep{ramirez2012, aguilera2018, bensbylind}, and Pop. II stars similar to HD 65907 conserve much more Li (\citealt{fuhrmann}, their Fig. 4). Furthermore, the combination of the stellar mass and metallicity indicates that it most likely does not belong to the Li dip (the sudden drop in Li abundances for a certain range of effective temperature or mass, usually observed in clusters; \citealt{li_dip}). According to Fig. 9 of \citet{francois}, the Li dip for a metallicity of [Fe/H] $\sim$ -0.3 dex occurs at a mass close to 1.1 M$_\odot$. The dip would fall near HD 65907's mass only for [Fe/H] $\sim$ -0.55 dex, which is significantly lower than the [Fe/H] determined for this star.

    The fact that the star has depleted such a considerable amount of its Li content, coupled with its $\alpha$ enhancement, [Y/Mg] ratio, and the issue with the isochronal age, suggests that HD 65907 suffered a mass accretion episode during which temperatures rose up above the Li burning temperature of $\sim 2.5 \times 10^6$ K. Therefore, this star most likely had a close companion that suffered from either a merger or a mass accretion phenomenon. In the latter case, the companion would still be detectable, as is the case in several white dwarf remnants in the literature (such as \citealt{schirbel}, \citealt{desidera}, and \citealt{fuhrmann17}, for radial velocity detections, or \citealt{crepp13}, \citealt{bacchus}, \citealt{crepp18}, and \citealt{gratton}, for direct detections via imaging).

    Another scenario proposed by \citet{fuhrmann} is that this star could have suffered a dynamical encounter with the progenitor cloud of the NGC 2516 cluster. This cluster is found within 30 arcmin of HD 65907, at a distance of 414 $\pm$ 9 pc \citep{ngc_spina}, and it has an age of $\sim$100 Myr. The star's and cluster's spatial distribution and kinematics provide some evidence that an encounter may have occurred. First, they share a similar V velocity component (rotational). Second, both HD 65907 and NGC 2516 are currently traveling toward the north Galactic pole, with NGC 2516 about 100 pc south of the Galactic plane (\citealt{fuhrmann}, their Fig. 6). Thus, it is also possible that the ``rebirth" of this star happened during an interaction with NGC 2516.

    One additional test to verify the blue straggler status of a star is to analyze its beryllium (Be) abundance \citep{schirbel, desidera}. Be is a less fragile element than Li (it is destroyed at T $\sim 3.5 \times 10^6$ K), and in solar twin stars its abundance is relatively constant with age \citep{tuccimaia}. This suggests that large Be depletion as a result of stellar evolution is unlikely, especially since HD 65907 is an F-type star that has a shallower convective zone than solar twins. Thus, if the star is Be-poor, it is likely a result of an interaction with a companion. A small Be depletion would be consistent with mass transfer, whereas a large Be depletion could only be explained by enhanced internal mixing due to angular momentum transfer or by a merger, during which temperatures would reach extreme values. A large Be depletion would also rule out the possibility that the rejuvenation of this star was due to an encounter with the progenitor cloud of NGC 2516.
        
    In this work we evaluate the Li and Be abundances of HD 65907 in a search for clues regarding what phenomena this star experienced. The chemical analysis is described in Sect. \ref{sec:abundances}. In Sect. \ref{sec:companions} we investigate time-series radial velocity data to look for a possible companion. We evaluate the photometric data and characterize a possible debris disk around this star in Sect. \ref{sec:disk}. The discussion of the results is presented in Sect. \ref{sec:discussion}, and our conclusions are summarized in Sect. \ref{sec:conclusions}.

\section{Li and Be abundances}
\label{sec:abundances}
The abundances of Li and Be were determined with the 2019 version of the 1D local thermodynamic equilibrium code \texttt{MOOG} \citep{moog}, adopting the \texttt{synth} driver. Given that both Li and Be are scarce elements, they need to be observed in their strongest spectral features, namely, their resonance lines. Since the Li resonance line (at 6707.8 \r{A}, in the red portion of the spectrum) and the Be resonance line (close to 3131 \r{A}, in the near-UV) fall in very different spectral regions, two different datasets were used. The abundances are reported as A(X) = log (N$_\text{X}$/N$_\text{H}$) + 12.

\subsection{Li}
The spectrum for the determination of A(Li) was composed by a combination of different spectra observed with the High Accuracy Radial velocity Planet Searcher (HARPS) spectrograph, taken from the European Southern Observatory (ESO) public database\footnote{ESO Science Archive Facility, \url{http://archive.eso.org/}.}. We queried for individual spectra with a signal-to-noise ratio (S/N) between 20 and 370, to avoid both noisy and saturated spectra, and aimed for a final spectrum with a S/N $\gtrsim$ 1000 close to the Li line.

The data reduction (Doppler correction, combination, and normalization) was performed using \texttt{IRAF}\footnote{Image Reduction and Analysis Facility, \url{https://iraf-community.github.io/}.} and information from the HARPS files. More details of the HARPS data processing can be found in \citet{eu}.

The input model atmosphere for \texttt{MOOG} was interpolated from the Kurucz model atmospheres ATLAS9 \citep{atlas9}, using the stellar parameters from \citet{shejeela}, who determined the effective temperature (5992 $\pm$ 9 K), log $g$ (4.52 $\pm$ 0.02 dex), [Fe/H] (-0.315 $\pm$ 0.005 dex) and the microturbulence velocity (1.24 $\pm$ 0.02 km/s) using HARPS data and the differential spectroscopic method (e.g., \citealt{bedell}; \citealt{melendez2014}). The macroturbulence velocity (3.78 km/s) and the projected rotational velocity ($v$ sin $i$ = 1.04 km/s) were calculated using the relations introduced by \citet{dossantos}. These relations require the full width at half maximum from the HARPS cross-correlation function; we adopted the average of the full width at half maximum values given for each observation.

To perform the spectral synthesis, the abundances of Li and other nearby atomic and molecular species are iteratively adjusted until the deviation from the observed spectrum is minimized. We adopted the line list from \citet{melendez2012}, which takes into account blends and the hyperfine structure of the Li line. Since the $^6$Li isotope is much less abundant than $^7$Li in the Sun \citep{asplund}, only the $^7$Li isotope was considered. 

\begin{figure}[!ht]
   \centering
   \includegraphics[width=\hsize]{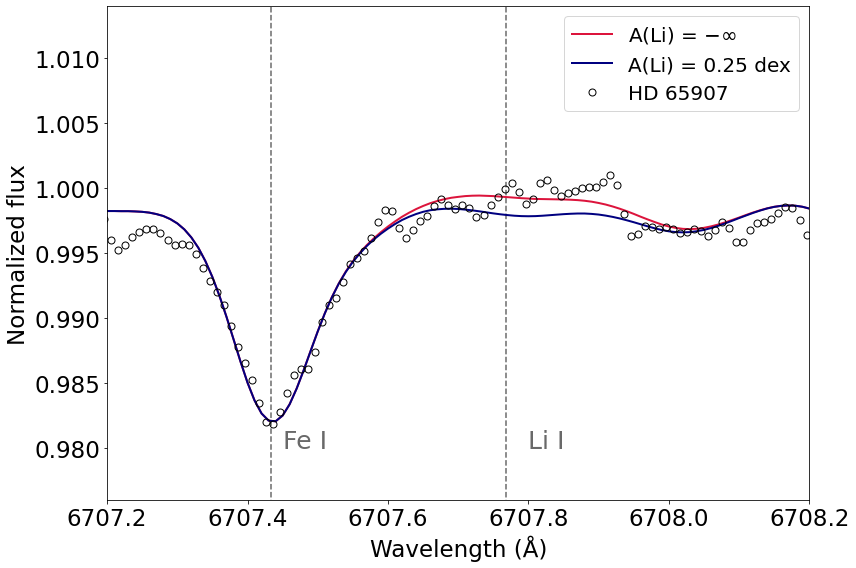}
      \caption{Spectral synthesis of the region of the $^7$Li resonance line for two different Li abundances (lines). The open circles represent the observed spectrum of the star HD 65907, with a S/N $\sim$1000.}
         \label{fig:li}
   \end{figure}

The comparison between the observed and the synthetic spectra in the region of the Li line can be seen in Fig. \ref{fig:li}. Despite the high resolving power of the HARPS data (R $\sim$115,000) and the high S/N of the combined spectrum ($\gtrsim$ 1000), only an upper limit of A(Li) $<$ 0.25 dex was detected. This result is in line with the low Li suggested by \citet{li4} (A(Li) $<$ 0.15 dex). Adopting a solar abundance of A(Li)$_\odot$ = 1.07 dex \citep{marilia}, the star HD 65907 is more Li-poor by at least a factor of 6. We note, however, that the data are still compatible with a total absence of Li in this star.

According to \citet[their Fig. 2]{ramirez2012}, stars with subsolar metallicity, solar mass, and T$_\text{eff} \sim$ 6000 K should have A(Li) between around 2.0 and 2.5 dex. This is similar to what was found by \citet{aguilera2018} for solar mass stars within the metallicity bin of -0.4 $\leq$ [Fe/H] (dex) $\leq$ -0.3 (their Fig. 7) or \citet{bensbylind} for stars with the mass, T$_\text{eff}$, and [Fe/H] of HD 65907 (their Fig. 5). Other works, such as \citet{gao} and \citet{giulia}, are also compatible with A(Li) $\sim$2 dex. This means that HD 65907 has at least 50 times less Li than similar stars, which provides clear evidence that this star underwent some phenomena that led to the destruction of almost all its Li content. 

\subsection{Be}
For the determination of A(Be), we adopted a Ultraviolet and Visual Echelle Spectrograph (UVES) spectra, again taken from ESO's public database. The S/N around the Be line is $\sim$20.

The process of data reduction was analogous to the case of the HARPS spectra, using the radial velocity available in the SIMBAD\footnote{SIMBAD Astronomical database, \url{http://simbad.cds.unistra.fr/simbad/}.} database. As explained in \citet{eu}, the HARPS spectra is split into seven parts before normalization. The UVES spectrum was not divided prior to the normalization, since it covers a much smaller spectral region (302 -- 388 nm vs. 378 -- 691 nm).

For the spectral synthesis, we adopted the same interpolated model atmosphere used for the Li synthesis and the line list from \citet{tuccimaia}, which includes atomic and molecular blends. To check the line list, we first performed the fit in a solar UVES spectrum (reflected in the asteroid Juno), adopting A(Be)$_\odot$ = 1.38 dex \citep{asplund}. The Be doublet resonance lines are found at $\lambda = 3130.420$ \r{A} (weaker) and $\lambda = 3131.065$ \r{A} (stronger). Since the stronger line is also more heavily blended, the final abundance was obtained based on the 3130.420 \r{A} fit, while the 3131.065 \r{A} fit was used for verification.

\begin{figure}[!ht]
   \centering
   \includegraphics[width=\hsize]{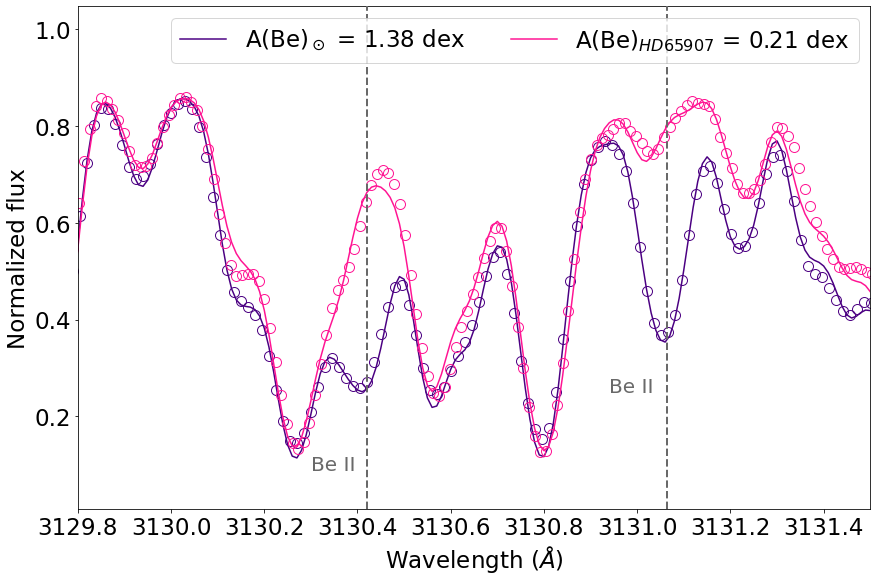}
      \caption{Spectral synthesis of the region of the Be resonance line for the Sun and the star HD 65907 (lines). The open circles represent the observed spectra.}
         \label{fig:be}
   \end{figure}

Figure \ref{fig:be} shows a comparison between the synthetic and observed spectra for the Sun and HD 65907. An upper limit of A(Be) $< 0.21$ dex was obtained for HD 65907, which means that the star has at least about 15 times less Be than the Sun.

As previously discussed, this excludes the need for invoking a dynamical encounter with the progenitor cloud of NGC 2516, as proposed by \citet{fuhrmann}. Such a severe Be depletion requires elevated temperatures, which can only be explained in the case of increased mixing due to angular momentum transfer from a companion or a stellar merger. This interaction with its binary pair would also be the cause of the low Li abundances and inconsistency between the chemical and isochronal age of HD 65907. The question that still needs to be answered is whether the mass accretion episode was due to mass transfer or a merger. In the case of mass transfer, the companion would still be detectable \citep{schirbel,desidera,fuhrmann17,gratton}. In the next section, we look for evidence of the presence or absence of this companion.


\section{Search for a companion}
\label{sec:companions}

It has long been known that the star HD 65907 is a member of a multiple system with three components \citep{mult_1997, mult_2008, mult_2017}. \citet{angular_sep} determined an angular separation from this star to the secondary of 60.6'' (corresponding to a projected separation of $\sim$983 AU; \citealt{mult_2016}), and from the secondary to the tertiary of 2.3''. Thus, any possible close interaction involving HD 65907 would have occurred with another (so far undetected) component.

To look for a signal of a close companion, we analyzed HARPS radial velocity data kindly provided by Gabriela Carvalho Silva via private communication. The dataset includes 112 data points spanning $\sim$14 years, after excluding low-quality data (spectra for which the S/N at the Ca H \& K lines was $<$ 80, the S-index deviated from the median by more than 3$\sigma$, or presented the unphysical negative flux flag from the code \texttt{ACTIN}\footnote{\texttt{ACTIN} code, \url{https://github.com/gomesdasilva/ACTIN}, \citet{gomes18}.}, as described by \citealt{gomes}, were filtered out).

\begin{figure}[!ht]
   \centering
   \includegraphics[width=\hsize]{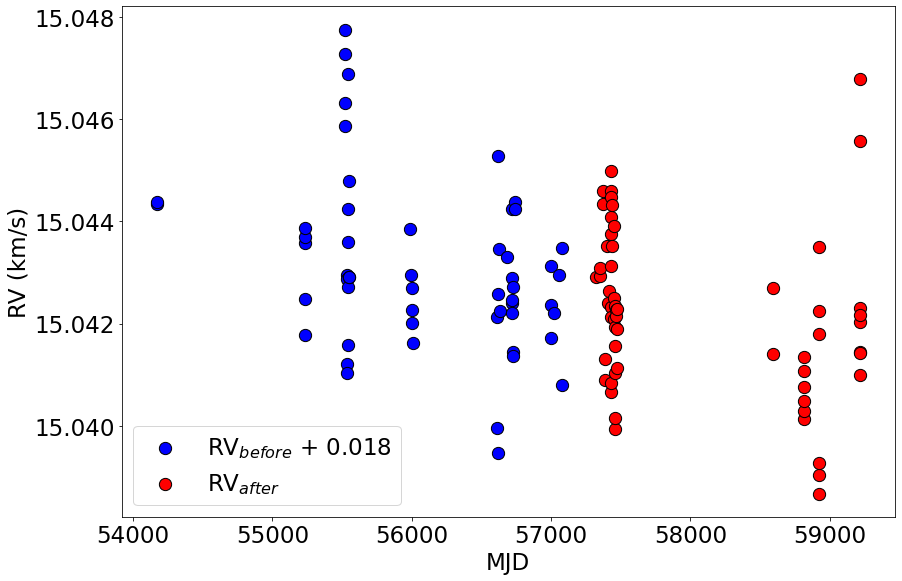}
      \caption{HARPS radial velocity curve for HD 65907. Radial velocity values from before the HARPS update were shifted by 0.018 km/s.}
         \label{fig:rv}
   \end{figure}

The resulting radial velocity time series is shown in Fig. \ref{fig:rv}. On June 3, 2015 (MJD 57176.5), the HARPS fiber was upgraded, and the radial velocity values from before and after this date presented a systematic difference. To correct for this effect, we added an offset of 0.018 km/s to the radial velocity values from before the upgrade. This offset was chosen based on the median radial velocities taken around 300 days before and after the upgrade (last blue and first red ``chunks" of data points in Fig. \ref{fig:rv}), ensuring continuity in the radial velocity curve.


\begin{figure}[!ht]
   \centering
   \includegraphics[width=\hsize]{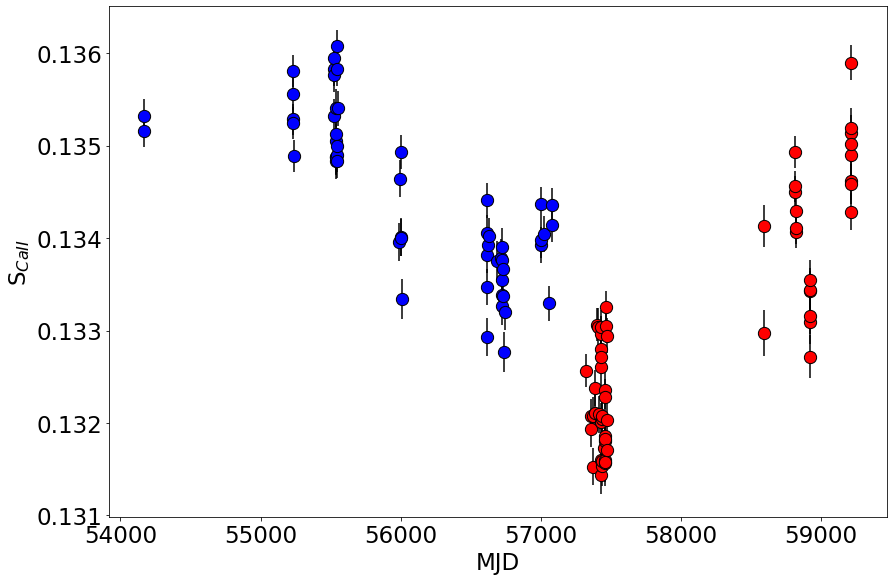}
      \caption{S-index from Ca II lines vs. the modified Julian date for the star HD 65907. Different colors indicate whether the data point was taken before (blue) or after (red) the HARPS upgrade.}
         \label{fig:s_index}
   \end{figure}

Figure \ref{fig:rv} shows a slight modulation in the radial velocity of about 2 m/s, with a downward trend at the beginning of the series and an upward trend at the end. We translated our $v$ sin $i$ into $P_\text{rot}$/sin $i$ and compared it with the stellar rotational period determined by \citet{prot} (13.8 days, based on TESS data) to find an inclination angle of $\sim$16°. Thus, the modulation in radial velocity corrected by sin $i$ is of $\sim$7.4 m/s, evidencing the lack of a companion -- the order of magnitude of the radial velocity variation in the presence of a companion should be much greater than what is observed. In any case, to investigate possible causes for the trends, we verified the variations in the stellar activity with time using the S-index from Ca II lines (provided by Gabriela Carvalho Silva via private communication, based on HARPS data). This is shown in Fig. \ref{fig:s_index}.

\begin{figure}[!ht]
   \centering
   \includegraphics[width=\hsize]{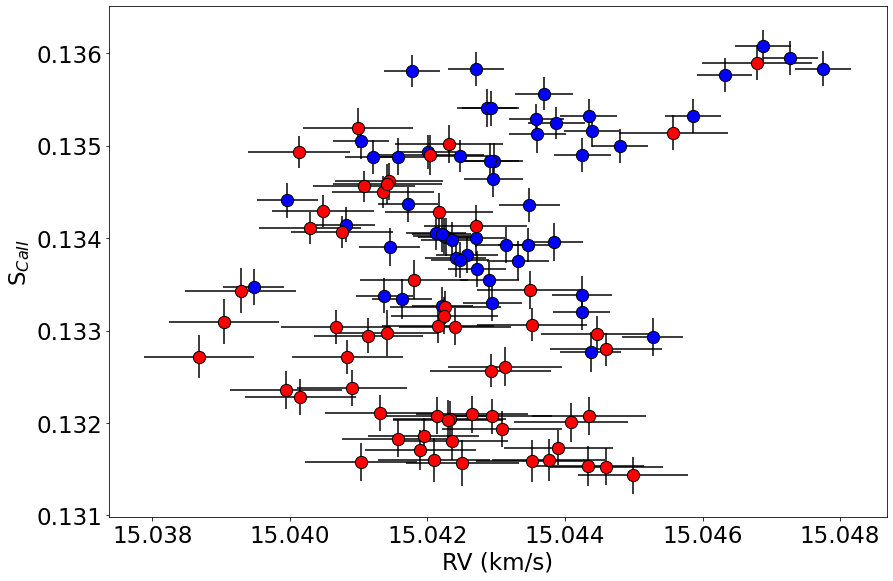}
      \caption{S-index vs. radial velocity for HD 65907. Radial velocity values from before the HARPS upgrade were shifted by 0.018 km/s. Different colors indicate whether the data point was taken before (blue) or after (red) the HARPS upgrade.}
         \label{fig:s_rv}
   \end{figure}

Figure \ref{fig:s_index} shows a modulation compatible with a stellar activity cycle. This modulation resembles what is seen in the radial velocity time series. In Fig. \ref{fig:s_rv} we plot the S-index versus radial velocity data for HD 65907. From Fig. \ref{fig:s_rv}, we can see that the radial velocity points with values a few m/s higher than the general distribution are associated with higher chromospheric activity levels in the star. To quantify this correlation, we performed two linear fits to this dataset, separating the points with downward or upward trends from Fig. \ref{fig:s_index} (which essentially meant separating points from before and after the HARPS upgrade, excluding the two outliers from after the upgrade that appear on the upper right part of Fig. \ref{fig:s_rv}). The fits show strong correlation between the two variables, with significances of $\sim$5.3 (before the upgrade) and $\sim$3.6 (after the upgrade). Therefore, the modulation in radial velocity is most likely caused by the stellar activity cycle, and we find no signs of a companion for HD 65907, which strongly favors the stellar merger scenario.

\section{Spectral energy distribution}
\label{sec:disk}

\begin{figure}[!ht]
   \centering
   \includegraphics[width=\hsize]{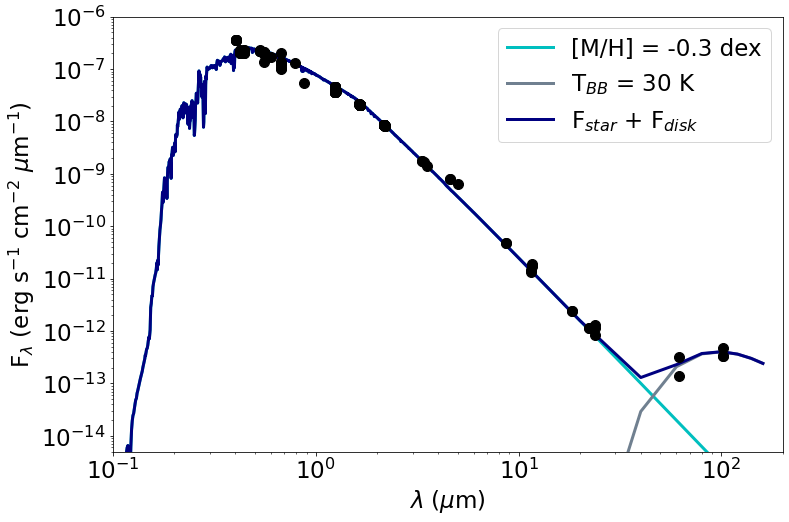}
   \caption{Spectral energy distribution for HD 65907. Lines indicate the interpolated Kurucz model for [M/H] = -0.3 dex (cyan), the scaled blackbody curve for a temperature of 30 K (gray), and the sum of the two (dark blue).}
         \label{fig:sed}
   \end{figure}

Previous works attempted to identify excess infrared emission from HD 65907. \citet{lawler} and \citet{sierchio} analyzed \textit{Spitzer} data and found no significant excess emission, whereas \citet{cruzalebes}, based on mid-IR data from multiple catalogs, noted an extended emission in the infrared, while still not classifying it as an excess. To update this investigation, we analyzed the spectral energy distribution of HD 65907 using photometric data available in the VizieR database\footnote{VizieR Catalog Service, \url{https://vizier.cds.unistra.fr/}.} with a search radius of 5'', and compared it with a theoretical model for a star with T$_\text{eff}$ = 6000 K, log $g$ = 4.5 dex, and metallicity [M/H] = -0.3 dex, interpolated from the Kurucz ATLAS9 models. This comparison revealed an excess infrared emission that was well reproduced by a 30 K blackbody model scaled to match the data. This is shown in Fig. \ref{fig:sed}.

This excess emission was interpreted as thermal emission from a debris disk surrounding this star \citep{trilling,liu,montesinos}. With the temperature for the disk ($T_d$, in K) found through blackbody modeling and the stellar luminosity ($L_\star$ = 1.26 $L_\odot$) determined via isochronal fitting (J. Shejeelammal, private communication), the orbital distance of the disk ($R_d$, in AU) can be found through Eq. (3) from \citet{ref_eq}:
\begin{equation}
    R_d = \left(\frac{278}{T_d}\right)^2 (L_\star)^{1/2}.
\end{equation}
For the debris disk of HD 65907, we find an orbital distance of $R_d = 96$ AU.

The detected debris disk should not be primordial, since the dust grains that compose debris disks usually have short lifetimes. However, a fraction of solar-type stars appear to maintain a debris disk for many years (\citealt{trilling}, their Fig. 13; \citealt{montesinos}); thus, this detection does not necessarily make this star a peculiar object.


\section{Discussion}
\label{sec:discussion}

As presented in Sect. \ref{sec:companions}, HD 65907 shows a highly stable radial velocity, with an average of 15.043 km/s ($\sigma = $ 0.002 km/s). The modulation in radial velocity, corrected by sin $i$, is of $\sim$7.4 m/s, which indicates the lack of a close companion. The few points that differ from the average radial velocity are associated with higher S-indexes, and are most likely the result of higher chromospheric activity levels. This argument is further strengthened considering the significances of the correlations found between the S-index and the radial velocity (Sect. \ref{sec:companions}).

The lack of evidence for a close companion based on radial velocity, in addition to the large destruction of Li and Be, reinforces the merger hypothesis -- the extreme temperatures reached by the objects during the merger would have been enough to deplete almost all (if not all) of their Li and most of their Be, which is compatible with the observed data. Therefore, the most likely scenario is that the star HD 65907 is a merger of two Pop. II star with ages of $\sim$11 Gyr. Its isochronal age of $\sim$5 Gyr is thus an indication of the amount of time that has passed since the merger. 

\begin{figure}[!ht]
   \centering
   \includegraphics[width=\hsize]{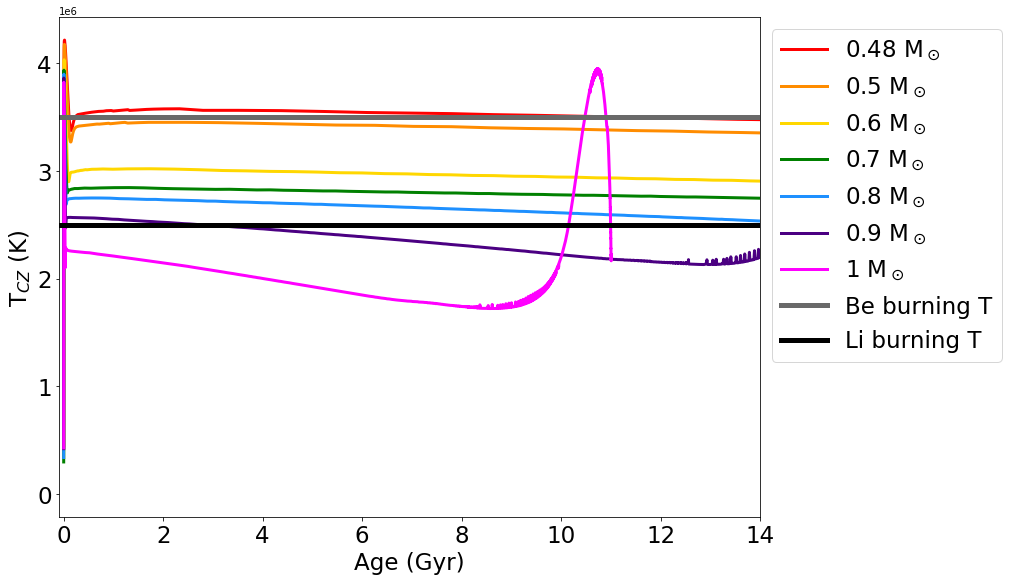}
   \caption{Temperature at the base of the convective zone vs. age for stars of different masses from the YaPSI stellar models. The temperatures needed for the destruction of Li and Be are indicated.}
         \label{fig:tcz}
   \end{figure}

Another possibility is that Li and Be depletion occurred before the merger, during the main sequence phase of the progenitor stars. However, from theoretical models (YaPSI; \citealt{spada}), shown in Fig. \ref{fig:tcz}, stars with masses $\gtrsim$ 0.5 M$_\odot$ and $\gtrsim$ 0.95 M$_\odot$ are able to conserve Be and Li, respectively. Hence, given the mass of HD 65907 (1.02$^{+0.01}_{-0.02}$ M$_\odot$; \citealt{shejeela}) the destruction of Be would only occur before the merger if the two progenitor stars had similar masses close to 0.5 M$_\odot$. If one of the merging stars was more massive than 0.5 M$_\odot$, then only the lower-mass binary would contribute with Be-depleted material, and the merger would still be required to explain the observed Be abundance of HD 65907.

From Fig. \ref{fig:tcz}, we can see that at the beginning of their lives, the convective region of all stars briefly reach high enough temperatures to destroy Li and Be. However, the short timescale of this temperature variation does not allow for much element depletion. This was observationally confirmed by \citet{rodolfo}, who determined Be abundances for low-mass stars (0.80 - 1.20 M$_\odot$) belonging to the two near zero age main sequence clusters IC 2391 and IC 2602 and found no difference in the Be content with effective temperature (their Fig. 6), which was compatible with theoretical predictions (their Figs. 10 and 12).

The mass of this blue straggler may seem low compared to the masses of other known blue stragglers (for example, \citealt{nemec} found an average mass of 1.3 M$_\odot$ for the blue stragglers in the globular cluster NGC 5466, and \citealt{ferraro97} determined $\sim$1.5 M$_\odot$ as the mean mass of blue stragglers in the globular cluster M3). On the other hand, HD 65907's mass is compatible with the mass range of blue stragglers in the globular cluster 47 Tuc (0.95 to 1.35 M$_\odot$, with a peak at 1.1 M$_\odot$; \citealt{ferraro2006}, their Fig. 3), and with the mass range of blue stragglers in open clusters (1 to 15 M$_\odot$, with the majority having masses $<$ 3 M$_\odot$; \citealt{jadhav}). Furthermore, the study by \citet{leinergeller} based on blue stragglers detected with \textit{Gaia} DR2 shows that the mass of blue stragglers in the oldest open clusters starts at about 1 solar mass (their Fig. 3). Other field blue stragglers also present masses around the solar value \citep{schirbel, gratton}.

It is interesting to note that $\sim$5 Gyr after the merger, the stellar rotational velocity (corrected for sin $i$) has already fallen to the low value of 3.85 km/s. This is not surprising -- most blue stragglers identified in clusters are slow rotators, as would be expected for their masses and ages \citep{sills16}. However, the exact behavior of the angular momentum following a stellar merger is still unclear. For blue stragglers that result from mass transfer, we observe an increase on the rotational velocities soon after the event, and their fast rotation can be used to identify blue straggler candidates (for example, \citealt{leiner} for blue stragglers in the M67 cluster). Additionally, there is evidence that the star spins down quite fast after such an event. \citet{leiner18} studied the relation between post-formation ages and rotation rates of mass transfer products and found that the stellar spin down can be described by magnetic braking prescriptions, much like normal solar-type stars. Therefore, it is possible that the evolution of the angular momentum happens similarly in all blue stragglers, regardless of how they were formed, and the low rotational velocity of HD 65907 would thus be expected, especially considering the amount of time ($\sim$5 Gyr) since the merger.

\begin{figure}[!ht]
   \centering
   \includegraphics[width=\hsize]{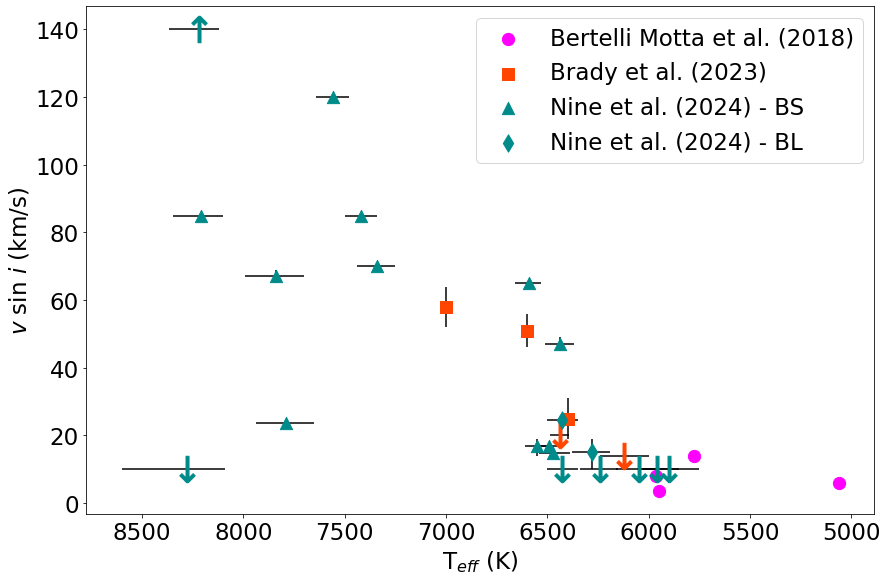}
   \caption{Projected rotational velocity vs. effective temperature for blue straggler stars in the M67 open cluster. The data were taken from \citet{bm}, \citet{brady}, and \citet{nine}. \citet{nine} separate the stars between blue stragglers (BS) and blue lurkers (BL), which are a lower-mass subset of blue straggler stars. Downward (upward) arrows indicate upper (lower) limits.}
         \label{fig:vsini}
   \end{figure}

In particular, HD 65907's projected rotational velocity is compatible to what is predicted for a star of its spectral type (F9.5V) -- these stars rotate much more slowly than earlier-type stars \citep{gray}. This effect of lower projected rotational velocities for lower-mass stars is also observed for blue stragglers in clusters. Figure \ref{fig:vsini} shows a collection of $v$ sin $i$ plotted against effective temperature for blue stragglers in the $\sim$solar age M67 cluster, and it shows that the blue stragglers with temperatures close to HD 65907's also have low rotational velocities. Furthermore, according to \citet{leiner}, the blue stragglers in M67 were formed less than 1 Gyr ago. Due to magnetic braking, their rotational velocities will decrease even further when they reach the same blue straggler age as HD 65907 ($\sim$5 Gyr).

The slightly larger-than-solar rotational velocity is also expected for a star with $\sim$solar mass and subsolar metallicity -- \citet{amard} show via simulations that, for the same mass, a lower [Fe/H] implies a lower rotational period, and thus a faster rotation rate (their Fig. 2). A star with lower metallicity will also spin down more slowly, and therefore it is expected that a star with half the solar [Fe/H], a mass close to solar, and a similar (post-merger) isochronal age will rotate faster than the Sun.

In addition to the low rotation, the current stellar activity levels are low (Fig. \ref{fig:s_index}). \citet{diego} and Gabriela Carvalho Silva (private communication) determined relationships between S-indexes and stellar ages for solar-type stars. The average chromospheric age for HD 65907, based on both relations, is 4.8 $\pm$ 1.3 Gyr, in reasonable agreement with the isochronal age. Thus, the stellar activity post-merger appears to have evolved as it would for a normal (non-blue straggler) solar-type star. The star still seems to maintain an activity cycle (shown in Fig. \ref{fig:s_index}) and a debris disk (Fig. \ref{fig:sed}), typical of young Pop. I stars. According to simulations performed by \citet{bluestrag_amanda}, the evolutionary tracks of blue stragglers following a stellar collision closely resemble their normal counterparts, but with a shorter main sequence lifetime. Although they do not discuss the evolution of blue stragglers that result from mergers, \citet{chen} argue that the merger and the collision products will present the same structure, indicating that the normal evolution of this star after the merger is not unexpected. Thus, the only trace of the HD 65907's past life lies in its chemical and kinematic signatures.

In the context of stellar multiplicity, the HD 65907 system is remarkably interesting -- it appears to have been originally a Pop. II quadruple system in which the inner binaries merged, producing a triple system with a blue straggler primary. The fact that the star was born in a multiple system is not surprising -- in a recent review, \citet{offner} found that about 14\% of solar-type stars are primaries in triple and higher order systems (their Fig. 1), and the excess of wide young stellar multiples in comparison with older field stars shows that most stars are born in multiple systems. However, it is known that external perturbations from Milky Way tides and passing field stars can disturb stellar systems, sometimes causing disruption or stellar collisions (e.g., \citealt{kaib}); thus, most systems do not survive their original configurations.

The survival probability of a quadruple system is not well defined, but the existence of such systems is not unheard of -- as an example, \citet{gliese} discovered a fourth component of the old thick disk stable system HD 40887. \citet{zhuchkov} analyzed via simulations the dynamical stability of the two hierarchical (and thus likely stable) quadruple systems, HD 68255/6/7 and HD 76644. While the former maintained its stability throughout the simulations, the evolution of the latter in many cases resulted in the merging of the two inner binaries. Therefore, our proposed evolutionary path for HD 65907 is reasonable.


\section{Conclusions}
\label{sec:conclusions}

In this work we have determined Li and Be abundances for the star HD 65907, whose chemical abundance and velocity components are evidence that its true age is $\sim$11 Gyr, despite its isochronal age of close to 5 Gyr \citep{fuhrmann, shejeela}. This provided further clues that this star underwent an episode of mass accretion  from a close companion. This episode could have been either mass and angular momentum transfer from an evolved companion, which would have increased rotational mixing and resulted in light-element depletion, or a stellar merger, during which the temperatures could have reached high enough values to result in severe Li and Be burning. We also studied its radial velocity time series, in a search for a possible companion, and analyzed the spectral energy distribution for this star.

Our conclusions can be summarized as follows:

   \begin{enumerate}
      \item The $^7$Li resonance line has a flat profile, consistent with complete Li depletion. An upper limit of A(Li) $<$ 0.25 dex is determined, which makes HD 65907 more Li-poor than the Sun by at least a factor of 6, and more Li-depleted than similar stars by at least a factor of 50.
      \item The star has a low Be abundance. We detect an upper limit of A(Be) $<$ 0.21 dex, making it at least 15 times more Be-depleted than the Sun.
      \item The radial velocity time series for HD 65907 shows little variation, about 2 m/s (or about 7.4 m/s when corrected for sin $i$), which is consistent with its stellar activity modulation. In particular, high radial velocity values are associated with high S-index values. This indicates that the star does not have a close companion and that the variations in radial velocity are a result of changes in the stellar activity.
      \item The above points provide strong evidence in favor of the merger scenario. Thus, HD 65907's true age is its chemical age ($\sim$11 Gyr; \citealt{shejeela}), and its isochronal age of $\sim$5 Gyr gives the time since the merger.
      \item As predicted by \citet{bluestrag_amanda} and \citet{chen}, the star has evolved normally since the merger. It has a low rotational velocity, its stellar activity is in agreement with what would be expected for a $\sim$5 Gyr solar-type star, and it still presents an activity cycle and a debris disk, which is not uncommon in other solar-type stars \citep{diego, trilling, montesinos}. 
      \item A blackbody fit to the infrared excess emission gives a temperature of 30 K for the debris disk. This locates the disk at a distance of 96 AU.
      \item The determination of the Be abundance was essential for discarding hypotheses concerning the the star's history, evidencing the importance of this element in the study of field blue stragglers (as proposed by \citealt{desidera}), which are much more difficult to confirm than blue stragglers belonging to stellar clusters. 
      
   \end{enumerate}

\begin{acknowledgements}
    AR thanks Fundação de Amparo à Pesquisa do Estado de São Paulo (FAPESP) for the Ph.D. funding (process no. 2023/07617-5). JM acknowledges the support from FAPESP (via process no. 2018/04055-8). AIK was supported by the Australian Research Council Centre of Excellence for All Sky Astrophysics in 3 Dimensions (ASTRO 3D), through project number CE170100013.

    We thank Dr. Rodolfo Smiljanic for the assistance with the UVES spectra.
    
    Based on observations collected at the European Southern Observatory under ESO programs 072.C-0488(E), 074.C-0364(A), 084.C-0229(A), 086.C-0230(A), 088.C-0011(A), 091.C-0936(A), 092.C-0579(A), 093.C-0062(A), 094.C-0797(A), 095.C-0040(A), 096.C-0053(A), 096.C-0499(A), 097.C-0021(A), 198.C-0836(A), 60.A-9700(G), 60.A-9709(G), and 093.D-0328(A).
\end{acknowledgements}

\bibliographystyle{aa}
\bibliography{bibliography}

\end{document}